\documentclass[article]{iopart}
\usepackage{iopams}
\usepackage[dvips]{color}

\begin{document}

\title[Holographic q-entropies]
{On R\'enyi entropy for free conformal fields: holographic and q-analog recipes}

\author{R Aros$^{\dag}$, F Bugini$^{\S}$ and D E D\'{\i}az$^{\dag}$}

\address{${\dag}$ Universidad Andres Bello,
Departamento de Ciencias Fisicas,
Republica 220, Santiago, Chile

${\S}$ Departamento de Fisica, Universidad de Concepcion, Casilla 160-C, Concepcion, Chile}

\ead{raros,danilodiaz@unab.cl, fbugini@udec.cl}

\begin{abstract}
  We describe a holographic approach to explicitly compute the universal logarithmic contributions to entanglement and R\'enyi entropies for free conformal scalar and spinor fields on even-dimensional spheres. This holographic derivation proceeds in two steps: first, following Casini and Huerta, a conformal map to thermal entropy in a hyperbolic geometry; then, identification of the hyperbolic geometry with the conformal boundary of a bulk hyperbolic space and use of an AdS/CFT holographic formula to compute the resulting functional determinant. We explicitly verify the connection with the type-A trace anomaly for the entanglement entropy, whereas the R\'enyi entropy is computed with aid of the Sommerfeld formula in order to deal with a conical defect.
  As a by-product, we show that the log-coefficient of the R\'enyi entropy for round spheres can be efficiently obtained as the q-analog of a procedure similar to the one found by Cappelli and D'Appollonio that rendered the type-A trace anomaly.
\end{abstract}



\section{Introduction}

Entanglement is certainly one of the most peculiar non-classical features of quantum physics, it entails correlations between separate systems which could never exist in any classical setting. Coming to grips with this novel concept led to the celebrated EPR~\cite{EPR} and Schr\"odinger's cat~\cite{Schr} paradoxes in the early years of quantum mechanics, and later on to Bell's inequalities~\cite{Bell} and their first experimental tests~\cite{Asp}. However, it was only in the last decade of the 20th century that quantum correlations started to be exploited as an information-theoretic resource in what nowadays has become the new and vast field of quantum information (see, e.g.~\cite{Horodecki:2009zz,Diosi}). Curiously, insights from quantum information theory, such as monogamy of entanglement, have fueled the renewed interest in the black hole information paradox of the last two years by means of novel `Gedankenexperimenten' (firewalls~\cite{Almheiri:2012rt}, wormholes~\cite{Maldacena:2013xja}, etc.).
\\
In the context of quantum field theory, on the other hand, entanglement and its associated entropy first turned up precisely in connection to the Bekenstein-Hawking area law of black hole entropy~\cite{'tHooft:1984re,Sorkin:2014kta,Bombelli:1986rw}. For a special conformal field theory (CFT), namely a free massless boson, Srednicki~\cite{Srednicki:1993im} showed that even in flat spacetime the entanglement entropy across a spherical surface is proportional to its area, but the coefficient is non-universal since it depends on the UV cut-off. This leading area term is a generic feature of CFTs (see, e.g.~\cite{Eisert:2008ur}) and there is a holographic prescription within AdS/CFT correspondence due to Ryu and Takayanagi~\cite{Ryu:2006bv} for CFTs with certain gravity duals.
\\
Our present interest, however, focuses on universal contributions to entanglement and R\'enyi entropy of free conformal fields. In even dimensions, these universal terms appear in logarithmic corrections (in the UV cut-off $\epsilon$) to the leading area law. The first indication occurs in two-dimensional CFTs, the entanglement entropy of an interval~\cite{Holzhey:1994we} is given by
\begin{equation}
S_{EE}=-\frac{c}{3}\,\log\epsilon~,
\end{equation}
with a coefficient proportional to the central charge $c$. This is the only overlap with the prescription by Ryu and Takayanagi, since in higher dimensions the universal logarithmic contributions are subleading with respect to the area term. Their generic structure for spherical entangling surfaces is fairly well understood by now (see, e.g.~\cite{Solodukhin:2011gn} and references therein), the coefficient $a$ of the type-A trace anomaly~\cite{Deser:1993yx} replaces the two-dimensional $c$. There are few explicit computations~\cite{Casini:2010kt,Lohmayer:2009sq,Dowker:2010nq,Dowker:2010bu,Solodukhin:2010pk} for the massless scalar, as well as general arguments~\cite{Ryu:2006ef,Myers:2010tj,Casini:2011kv} supporting this connection to the trace anomaly. 
R\'enyi entropy, despite its richness as a generalization (in fact, a $q-$deformation) of Shannon or von-Neumann measures of entropy, has only appeared in this QFT context mainly as an auxiliary tool in the computation of entanglement entropy (limiting value $q=1$). The $q$ parameter enters the computation by Casini and Huerta~\cite{Casini:2010kt} as length of a temperature circle; Dowker~\cite{Dowker:2010nq,Dowker:2010bu} considers the trace anomaly in the d-lune $S^d/\mathbb{Z}_q$ whereas the approach of Klebanov et al.~\cite{Klebanov:2011uf} involves $q-$fold coverings of the d-sphere; Solodukhin~\cite{Solodukhin:2010pk} introduces $q$ as a conical defect at the horizon of an extremal black hole; and finally, in the holographic computations of Hung et al.~\cite{Hung:2011nu}, $q$ enters in the temperature of a bulk topological black hole.
The R\'enyi entropy of an interval in two-dimensional CFTs~\cite{Calabrese:2004eu} is simply given by 
\begin{equation}
S_{q}=-\frac{c}{6}\,(1+\frac{1}{q})\,\log\epsilon~,
\end{equation} 
with the central charge $c$ intact. For free conformal fields and spherical entangling surfaces in higher even dimensions, however, the coefficient of the universal logarithmic terms has a polynomial dependence in $q^{-1}$ that only in the limit $q\rightarrow 1$ reduces to the type-A anomaly coefficient (see, e.g.~\cite{Lee:2014xwa}).\\
In this paper we present a holographic derivation of the universal logarithmic contributions to the R\'enyi entropy of free massless scalar and Dirac fields on even-dimensional spheres. 
We start with the entanglement entropy in section 2 as a warm up. In section 3 we show how these holographic results can be adapted to the case of R\'enyi entropy using heat kernel techniques. We find, in passing, a neat q-recipe to compute the polynomials in the deformation parameter. In section 4, we extend all these findings to the Dirac field. Concluding remarks are given in section 5.
\section{Entanglement entropy and type-A trace anomaly}

The main breakthrough in the analytic calculation of the vacuum entanglement entropy of a massless scalar on spheres in flat spacetime was achieved by the mapping to a thermal gas of free conformal scalars at temperature `$2\pi$' in the product geometry $R\times H^{d-1}$, in a manner familiar from the Unruh effect~\footnote{We thank Michael Stiller for information on this point and its relation with Algebraic Quantum Field Theory.}. This was first realized in~\cite{Casini:2010kt}, further elaborated in~\cite{Casini:2011kv} and put to work in, for example,~\cite{Klebanov:2011uf,Hung:2011nu,Huang:2014gca}.

\subsection{Thermal entropy}

Here we briefly reproduce the direct heat-kernel calculation from~\cite{Casini:2010kt}. The partition function can be first computed from the functional determinant for the conformal Laplacian $Y$ in the geometry $S^1(\beta)\times H^{d-1}$ and then one can read off the entropy.

\begin{equation}
Z(\beta)=\exp\{-W(\beta)\}=[\,\det Y \,]^{-1/2}.
\end{equation}

\begin{equation}
S_{EE}=[\beta\partial _{\beta}-1]\,W(\beta)\mid _{\beta=2\pi} .
\end{equation}
The effective action is obtained from the diagonal heat kernel

\begin{equation}
W(\beta)=-\frac{1}{2}\tr \int_0^{\infty} \frac{dt}{t}\, K_{S^1}(t)\times K_{H^{d-1}}(t).
\end{equation}
Notice that terms linear in $\beta$ do not contribute to the entropy, so that the useful information is contained in contributions from the images on the temperature circle

\begin{equation}
K'_{S^1}(t)=\frac{2}{\sqrt{4\pi t}}\sum_{m=1}^{\infty} e^{-\frac{m^2 \beta^2}{4t}}.
\end{equation}
There are no image contributions for the hyperbolic space, being noncompact, and the heat kernel has only finite non-vanishing terms

\begin{equation}
K_{H^{d-1}}(t)=\frac{1}{(4\pi t)^{d/2-1/2}}\sum_{n=0}^{d/2-2} a_n^{(d-1)}\,t^n.
\end{equation}
The infinite volume has to be taken into account and the logarithmic term ${\cal L}_{d-1}=\frac{2(-\pi)^{d/2-1}}{\Gamma(d/2)}$ comes precisely from this volume regularization~\cite{Gra99,Albin:2005}.
The outcome, after proper time integration, is an explicit expression in terms of the heat-kernel coefficients of the conformal Laplacian in hyperbolic space and Riemann zetas

\begin{equation}
\label{CH10}
g_0=\frac{(-1)^{d/2}}{2^{d-3}\Gamma(d/2)}\sum_{n=0}^{d/2-2} \,a_n^{(d-1)}\,\pi^{2n-d}\,\Gamma(d/2+1-n)\,\zeta[d-2n]~.
\end{equation}
This is the result obtained by Casini and Huerta~\cite{Casini:2010kt} when the sum in their overall coefficient is worked out in detail.

\subsection{Holographic approach}

The holographic derivation we follow in this case of entanglement entropy is to compute the partition function on the d-sphere, conformal compactification of the thermal geometry ,  by use of an special entry in AdS/CFT dictionary, namely relating one-loop determinants for bulk fields in asymptotically AdS backgrounds and
determinants of correlation functions of the dual operators at the boundary~\cite{Gubser:2002zh,Gubser:2002vv,Hartman:2006dy,Diaz:2007an}.
Particular values of the scaling dimensions $\lambda=d/2 + k$ of the operators produce the conformal powers of the Laplacian or GJMS operators $P_{_{2k}}$ on the boundary and the conformal Laplacian $Y$ corresponds to the first ($k=1$) representative of this family.

\subsubsection{Green function:}

This holographic path was pursued in~\cite{Diaz:2008hy}
\begin{equation}
-\frac{1}{2}\log \det P_{_{2k}}\mid _{_{S^d}}\,=\, vol(H^{d+1})\cdot\int_0^k d\nu\,2\nu{\cal A}_d(\nu)~,
\end{equation}
where the polynomial in the integrand is essentially the Plancherel measure at imaginary argument
\begin{equation}
\label{CompactFormula}
2\nu{\cal A}_d(\nu)=\frac{1}{(4\pi)^{d/2}}\,\frac{(\nu)_{d/2}\cdot(-\nu)_{d/2}}{(1/2)_{d/2}}~,
\end{equation}
in terms of Pochhammer symbols.
This is in fact an equality if taken in the sense of dimensional regularization (DR), the logarithmic divergent term is produced by the volume anomaly (pole in DR) and the finite part renders the regularized determinant. The type-A trace anomaly coefficients obtained in this way were independently derived using spectral methods by Dowker~\cite{Dowker:2010qy}~\footnote{The factorization of the GJMS operators on the sphere was crucial in~\cite{Dowker:2010qy}; a couple of numerical missprints in~\cite{Diaz:2008hy} were detected, but the same compact formula was obtained.}.

\subsubsection{Heat kernel:}

Alternatively, one can compute with the heat kernel in $H^{d+1}$ as follows 

\begin{equation}
\tr G_+=\tr \frac{1}{-\nabla^2+m_k^2}=\tr \int_0^{\infty} \frac{dt}{t} \,e^{\nabla^2t-m_k^2t}~,
\end{equation}
with resonant values for the mass $m_k^2=d^2/4+k^2$.
The diagonal heat kernel in hyperbolic space has a finite expansion

\begin{equation}
\tr e^{\nabla^2t-m_k^2t}\,=\,vol(H^{d+1})\cdot\frac{e^{-\nu^2t}}{(4\pi)^{d/2+1/2}} \sum_{n=0}^{d/2-1} \,a_n^{(d+1)}\,t^n~.
\end{equation}
After proper time integration, the answer in DR is finite and the $-$branch is obtained by analytic continuation in $\nu$. There is in the end an alternative expression for the trace anomaly coefficient, the polynomial in $k$ in terms of the heat kernel coefficients
\begin{equation}\fl
-\frac{1}{2}\log \det P_{_{2k}}\mid _{_{S^d}}\,=\, \frac{-vol(H^{d+1})}{(4\pi)^{d/2+1/2}}
\sum_{n=0}^{d/2-1} \,a_n^{(d+1)}\,\Gamma(n-1/2-d/2)\,k^{d+1-2n}~.
\end{equation}
This time the resulting expression only accounts for the anomaly terms and not for the regularized determinant.

\subsection{A recipe}

The aim now is to reproduce the exact combination of terms in the entanglement entropy (equation~\ref{CH10}) by manipulation of the previous expression for the type-A trace anomaly of GJMS operators. First, use reflection formula for the Riemann zeta~\footnote{This is equivalent to use Poisson re-summation in the heat kernel on the temperature circle.} to rewrite the result by Casini and Huerta as

\begin{equation}\fl
g_0\,=\, \frac{-vol(H^{d-1})}{(4\pi)^{d/2-1/2}}
\sum_{n=0}^{d/2-2} \,a_n^{(d-1)}\,(d-2n)\,\Gamma(n+1/2-d/2)\,\zeta[2n+1-d]~.
\end{equation}
This expression can be obtained by the following manipulation of the type-A trace anomaly coefficient of the GJMS operators
\begin{equation}\fl
-\frac{1}{2}\log \det P_{_{2k}}\mid _{_{S^{(d-2)}}}\,=\, \frac{-vol(H^{d-1})}{(4\pi)^{d/2-1/2}}
\sum_{n=0}^{d/2-2} \,a_n^{(d-1)}\,\Gamma(n+1/2-d/2)\,k^{d-2n-1}~.
\end{equation}
Multiply by $k$, then take the derivative with respect to $k$ to get the extra factor $(d-2n)$, and then replace the powers of $k$ by Riemann zetas evaluated at minus the corresponding power. The trick now is to perform these manipulations not to the expression in terms of heat kernel coefficients but rather to the compact formula (equation~\ref{CompactFormula}). Collecting all numerical coefficients, we end up with the following recipe for $g_0$~\footnote{We have intentionally traded $k$ for $z$ to make the recipe more mnemonic.}:
\begin{enumerate}
\item first compute the polynomial $$-2\frac{(-1)^{d/2}}{(d-2)!}\,\frac{d}{dz}\left\{ z\cdot\int_0^z d\nu\; (\nu)_{\frac{d}{2}-1}\cdot(-\nu)_{\frac{d}{2}-1}\right\}~,$$
\item then replace powers of zeta by Riemann zeta functions: $$z^l \mapsto \zeta[-l]~.$$
\end{enumerate}
This recipe is very reminiscent of that obtained by Cappelli and D'Appollonio~\cite{Cappelli:2000fe} to compute type-A anomaly coefficient of the conformal Laplacian, with the diefference that ours involves Riemann zeta functions whereas their recipe manipulates Bernoulli numbers.

\section{R\'enyi entropy}

After mapping to thermal entropy, the R\'enyi entropy can be computed from the knowledge of the thermal partition function $Z(\beta)$ at inverse temperature $2\pi q$ as follows
\begin{equation}
S_q=\frac{\log Z(2\pi q)-q\,\log Z(2\pi)}{1-q}~.
\end{equation}
Again, the logarithmic term comes the volume regularization of the spatial hyperbolic section. The overall coefficient $g_0^{q}$, polynomial in $q^{-1}$, is obtained as an explicit expression in terms of the heat-kernel coefficients of the conformal Laplacian in hyperbolic space and Riemann zetas~\cite{Casini:2010kt}.

\subsection{Holographic formula}

Let us now see how can we adapt our holographic derivation in order to account for the R\'enyi entropy.
The bulk interior $H_{*}^{d+1}$ is given by the following metric 
\begin{equation}
ds^2_{H_{*}^{d+1}}=d\mu^2+ \mbox{sh}^2\mu d\,\tau^2+ \mbox{ch}^2\mu \,ds^2_{H^{d-1}}~, 
\end{equation}
so that the conformal infinity at $\mu\rightarrow\infty$ is precisely the product geometry $S^1\times H^{d-1}$. Clearly, when $\tau$ is $2\pi-$periodic we have regular hyperbolic space; otherwise $\tau\equiv\tau+2\pi q$ produces an angular deficit at the deep interior. 
The plan is then to work out the heat kernel by means of Sommerfeld formula. The volume integral inherits the log-divergence of the (d-1)-dimensional hyperbolic section and it turns out that the universal coefficient is given essentially by a residue.
For illustration, we highlight the computation in two dimensions (d=2).\\
The heat kernel for the scalar is
\begin{equation}
K_3(\sigma;t)=\frac{e^{-t}}{2\pi} (-\frac{\partial}{\partial\mbox{ch}\sigma})\,\frac{e^{-\frac{\sigma^2}{4s}}}{\sqrt{4\pi s}}~;
\end{equation}
it only depends on the proper time $t$ and on the geodesic distance between $\sigma$ image points, with 
\begin{equation}
\mbox{sh}\frac{\sigma}{2}=\sin \frac{w}{2} \cdot \mbox{sh}\mu~,
\end{equation}
where $w$ is the angular separation along $\tau$. After change of variable in the volume integral from the radial $\mu$ to $\mbox{ch}\sigma$, the only term that enters in Sommerfeld's formula is $\frac{1}{\sin^2 \frac{w}{2}}$. Apart from numerical factors, the $q-$dependence of the R\'enyi entropy is read off from the following residue
\begin{equation}
\mbox{Res}\{\frac{\cot\frac{w}{2q}}{\sin^2\frac{w}{2}},w=0\}\,\sim\,(\frac{1}{q}-q)~.
\end{equation}
Finally, the quotient by $(1-q)$ renders the expected universal term $(1+\frac{1}{q})$.\\
A recursion is easily implemented for the higher dimensional cases and the results agree with those reported in the literature.

\subsection{A q-analog recipe}

For practical purposes, however, it is much more efficient the following procedure that we found out by examining in detail the modification implied by R\'enyi entropy to the recipe we had previously found for the entanglement entropy. The only necessary modification turns out to be the substitution of the ordinary derivative with respect to $z$ by the `$q^{-1}-$derivative'.
The $q-$derivative of a function $f$ is defined for $q\neq 1$ by 
\begin{equation}
\left[\frac{df}{dz}\right]_{q}=\frac{f(qx)-f(x)}{qx-x}~,
\end{equation} 
and it reduces to the ordinary derivative in the limit $q\rightarrow 1$.\\
In all, the recipe we get is the following
\begin{enumerate}
\item first compute the polynomial $$-2\frac{(-1)^{d/2}}{(d-2)!}\,\left[\frac{d}{dz}\right]_{q^{-1}}\left\{ z\cdot\int_0^z d\nu\; (\nu)_{\frac{d}{2}-1}\cdot(-\nu)_{\frac{d}{2}-1}\right\}~,$$
\item then replace powers of zeta by Riemann zeta functions: $$z^l \mapsto \zeta[-l]~.$$
\end{enumerate}

\section{Dirac spinor}

The massless Dirac spinor can be handled similarly. We just quote the compact formula for entanglement entropy~\cite{Aros:2011iz,Dowker:2013mba} 
\begin{equation}
g_0=\frac{(-1)^{\frac{d}{2}}\,2^{2+\frac{d}{2}}}{d!}\,\int_0^{\frac{1}{2}} d\nu\; (\frac{1}{2}+\nu)_{\frac{d}{2}}\cdot(\frac{1}{2}-\nu)_{\frac{d}{2}}\;~,
\end{equation}
and report the q-recipe for R\'enyi entropy
\begin{enumerate}
\item first compute the polynomial $$\frac{(-1)^{\frac{d}{2}}2^{1+\frac{d}{2}}}{(d-2)!}\,\left[\frac{d}{dz}\right]_{q^{-1}}\left\{ z\cdot\int_0^z d\nu\; ((\frac{1}{2}\nu)_{\frac{d}{2}-1}\cdot(\frac{1}{2}-\nu)_{\frac{d}{2}-1}\right\}~,$$
\item then replace powers of zeta by Hurwitz zeta functions: $$z^l \mapsto \zeta[-l,\frac{1}{2}]~.$$
\end{enumerate}

\section{Conclusion}

To summarize, we have found a holographic derivation of the universal logarithmic contributions to entanglement and R\'enyi entropies for free conformal scalar and spinor fields on even-dimensional spheres and efficient q-recipes in the case of R\'enyi entropy. The q-recipes bear a striking similarity, that we believe deserves further examination, with the remark due to Baez~\cite{Baez} that the R\'enyi entropy of a system in thermal equilibrium is minus the `$q^{-1}-$derivative' of its free energy with respect to temperature.\\
For type-A trace anomaly, the holographic and spectral approaches discussed here have recently been extended to conformal higher spin fields in a series of seminal papers~\cite{Giombi:2013yva,Tseytlin:2013jya,Tseytlin:2013fca}.
However, in connection with entanglement entropy, the controversy already existing for the gauge vector (see, e.g.~\cite{Eling:2013aqa}) will probably carry over to the whole tower of higher-spin gauge fields and a successful resolution would be much appreciated.

\ack This work was partially funded by grants FONDECYT 11110430, 1131075 and
UNAB DI-286-13/R, DI-295-13/R.



\section*{References}

\providecommand{\href}[2]{#2}\begingroup\raggedright\endgroup
\end{document}